\documentclass[aps,pra,twocolumn,floatfix,nofootinbib]{revtex4}
\usepackage{amssymb}
\usepackage{bm}
\usepackage{epsfig}

\begin{document}

\title{Oscillating axion bubbles as alternative to supermassive black holes
at galactic centers }
\author{Anatoly A. Svidzinsky}
\affiliation{Department of Physics, Institute for Quantum Studies, Texas A\&M University,
College Station, TX 77843-4242 \\ asvid@jewel.tamu.edu}
\date{\today }

\begin{abstract}
Recent observations of near-infrared and X-ray flares from Sagittarius A$%
^{\ast }$, which is believed to be a supermassive black hole at the Galactic
center, show that the source exhibits about 20-minute periodic variability.
Here we provide arguments based on a quantitative analysis that supermassive
objects at galactic centers may be bubbles of dark matter axions rather than
black holes. An oscillating axion bubble can explain periodic variability of
Sagittarius A$^{\ast }$ and yields the axion mass about $0.6$ meV which fits
in the open axion mass window. The bubble scenario with no other free
parameters explains lack of supermassive ``black holes" with mass $M<10^{6}$M%
$_{\odot }$. Low-mass bubbles decay fast and as a result are very rare. We
also found that the mass of an axion bubble can not exceed $1.5\times 10^{9}$%
M$_{\odot }$, in agreement with the upper limit on the supermassive ``black
hole" mass obtained from observations. Our finding, if confirmed, suggests
that Einstein general relativity is invalid for strong gravity and the
gravitational field for the bubble effectively becomes repulsive at large
potential. Imaging a shadow of the ``black hole" at the Galactic center with
VLBI in the next decade can distinguish between the black hole and the
oscillating axion bubble scenarios. In the case of axion bubble, a steady
shadow will not be observed. Instead, the shadow will appear and disappear
periodically with a period of about $20$ min.
\end{abstract}

\maketitle

\section{Introduction}

Originally introduced to explain why the strong interaction, in contrast to
weak interactions, does not violate CP symmetry \cite{Pecc77}, hypothetical
axions have since become one of the leading particle candidates for the cold
dark matter in the Universe \cite{Brad03}. The axion appears as a pseudo
Nambu-Goldstone boson of a spontaneously broken Peccei-Quinn symmetry \cite%
{Pecc77}, whose scale $f$ determines the axion mass $m$,%
\begin{equation}
m\approx \frac{m_{\pi }f_{\pi }}{2f}=0.62\text{ eV}\cdot \frac{10^{7}\text{
GeV}}{f}  \label{f1}
\end{equation}%
and suppresses the coupling to Standard Model particles, $\propto 1/f$. Here 
$m_{\pi }=135$ MeV is the neutral pion mass and $f_{\pi }=93$ MeV its decay
constant \cite{Raff02,Brad03}. Astrophysical and cosmological arguments
constrain the axion mass $m$ to be in the range of $10^{-6}-3\times 10^{-3}$
eV \cite{Brad03}. Axions in this mass range could provide much or all of the
cold dark matter in the Universe. Properties of stars impose the upper limit
on the axion mass via constraints on axion interaction with photons, leptons
and nucleons. However, such interactions are model-dependent. Observations
of the cosmological large-scale structure constrain the axion-pion coupling
which yield weaker upper mass limit $m<1.05$ eV \cite{Hann05}.

Interaction of axions with QCD instantons generates the axion mass and
periodic interaction potential \cite{Kim87} 
\begin{equation}
V(\varphi )=m^{2}f^{2}[1-\cos (\varphi /f)],  \label{p01}
\end{equation}%
where $\varphi $ is a real scalar axion field.

\begin{figure}[tbp]
\bigskip 
\centerline{\epsfxsize=0.45\textwidth\epsfysize=0.41\textwidth
\epsfbox{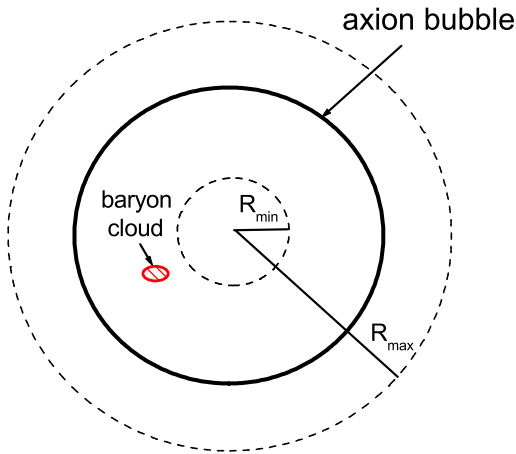}}
\caption{Mechanism of flare variability. Axion bubble radius oscillates
between $R_{\min }$ and $R_{\max }$ which yields periodic variation of the
gravitational redshift $z(t)$ of the bubble interior. The dark matter bubble
itself does not radiate electromagnetic waves; radiation is produced by
flaring clouds of baryon matter trapped in the bubble interior. Redshift $%
z(t)$ reduces radiation power of the cloud by a factor of $1/(1+z(t))^{2}$.
As a result, intensity of the flaring cloud radiation is modulated with the
period of bubble oscillation. Modulation amplitude depends on the distance
between the cloud and the bubble center. }
\label{u0}
\end{figure}

Here we argue that oscillating axion bubbles, rather then supermassive black
holes, could be present at galactic centers. Recent observations of
near-infrared and X-ray flares from Sagittarius A$^{\ast }$, which is
believed to be a $3.6\times 10^{6}$M$_{\odot }$ black hole at the Galactic
center, show that the source exhibits about 20-minute periodic variability 
\cite{Genz03,Gill06,Bela06}. An oscillating axion bubble can explain
periodic variability of Sagittarius A* and yields the axion mass about $0.6$
meV. Fig. \ref{u0} explains the mechanism of flare variability.

Moreover, the bubble scenario explains observed lack of supermassive
\textquotedblleft black holes" with mass $M<10^{6}$M$_{\odot }$. As we
discuss later in this paper the bubble life-time $t\propto M^{9/2}$, it
becomes less then the age of the Universe for $M\lesssim 5\times 10^{6}$M$%
_{\odot }$. The bubble at our Galactic center would decay within about $%
5\times 10^{9}$ yrs. If, however, $M<10^{6}$M$_{\odot }$ the decay time
becomes very short, $t\lesssim 10^{7}$ yrs, and as a result such objects are
very rare.

Finally, the axion bubbles with no free parameters (if we fix $m=0.6$ meV
based on Sagittarius A* flare variability) explain the upper limit ($%
1.5\times 10^{9}$M$_{\odot }$) on the supermassive \textquotedblleft black
hole" mass found in recent analysis of the measured mass distribution \cite%
{Wu02}.

In recent years, the evidence for the existence of an ultra-compact
concentration of dark mass associated with the radio source Sagittarius A*
in the Galactic Center has become very strong. However, an unambiguous proof
that this object is a black hole is still lacking. A defining characteristic
of a black hole is the event horizon. To a distant observer, the event
horizon casts a relatively large \textquotedblleft shadow\textquotedblright\
with an apparent diameter of about $10$ gravitational radii due to bending
of light. The predicted size ($\sim $30 micro-arcseconds) of this shadow for
Sagittarius A* approaches the resolution of current radio-interferometers.
Hence, there exists a realistic expectation of imaging the shadow of a black
hole with very long-baseline interferometry within the next decade \cite%
{Falc00,Falc01,Shen05,Huan07}. Such imaging will allow us to distinguish
between the black hole and the oscillating axion bubble scenario which we
propose in this paper. If the axion bubble, rather then a black hole, is
present at the Galactic center, the steady shadow will not be observed.
Instead, the shadow will appear and disappear periodically with a period of
about $20$ $\min $.

\section{Axion bubbles}

We introduce dimensionless coordinates and define the unit of distance, time
and $\varphi $ as 
\begin{equation}
r_{0}=\frac{\hbar }{mc},\quad t_{0}=\frac{\hbar }{mc^{2}},\quad \varphi _{0}=%
\frac{1}{\sqrt{4\pi G}},
\end{equation}%
where $c$ is the speed of light and $G$ is the gravitational constant. For
the moment we omit gravity. Further we use natural units for which $\hbar
=c=1$. Energy of the axion field in units of $m_{\text{pl}}^{2}/m$ is given
by 
\begin{equation}
E=\int d\mathbf{r}\left[ \frac{1}{2}\left( \frac{\partial \varphi }{\partial
t}\right) ^{2}+\frac{1}{2}(\nabla \varphi )^{2}+V(\varphi )\right] ,
\label{ef}
\end{equation}%
where 
\begin{equation}
V(\varphi )=\frac{1}{\alpha ^{2}}[1-\cos (\alpha \varphi )],\quad \alpha =%
\frac{1}{\sqrt{4\pi G}f}=\frac{m_{\text{pl}}}{\sqrt{4\pi }f}
\end{equation}%
is the dimensionless potential and the coupling parameter respectively, $m_{%
\text{pl}}=\sqrt{\hbar c/G}=1.2\times 10^{19}$ GeV is the Planck mass.

The interaction potential $V$ has degenerate minima $V=0$ at $\varphi =2\pi
n/\alpha $, where $n$ is an integer number. As a consequence, equation for
the axion field $\varphi $ has bubble-like solutions. The bubble surface is
an interface between two degenerate vacuum states with $\varphi =2\pi
n/\alpha $ ($r<R$) and $\varphi =0$ ($r>R$). In this paper we consider
spherical bubbles with surface width much smaller then its radius $R$ and $%
n=1$. Energy density of the axion field is nonzero only at the bubble
surface. Energy of a static thin-wall bubble is $E=4\pi \sigma R^{2}$, where 
$\sigma $ is the surface energy per unit area (which equals to the surface
tension for the domain wall we consider \cite{Ipse84}) determined by an
integral over one potential period \cite{Svid04,Svid04b} 
\begin{equation}
\sigma =\frac{1}{4\pi }\int \sqrt{2V}d\varphi =\frac{2}{\pi \alpha ^{2}}.
\label{y0}
\end{equation}%
Surface tension $\sigma $ depends only on the axion interaction strength.
The later, however, slightly depends on the axion model which can change Eq.
(\ref{y0}) by a factor of the order of one \cite{Huan85}.

Under the influence of surface tension an initially static bubble collapses
to its center which yields reduction of the surface energy. However, if we
include gravity this gives an additional energy contribution. Such a
contribution could substantially alter the bubble evolution and, in
particular, prevent collapse as we discuss in the next section.

One should mention that decay of axions into photons is suppressed in the
bubble. Such a decay is not allowed by energy conservation. The bubble
surface is approximately a one dimensional kink. In the kink's reference
frame the kink is static and the distribution of the axion field is obtained
by minimization of the energy functional (\ref{ef}) subject to the boundary
conditions that far from the kink we have fixed vacuum states ($\varphi =0$
from one side and $\varphi =2\pi /\alpha $ from the opposite side). The
optimized field distribution determines the kink's energy. Any small change
in $\varphi $ would increase the total energy of the axion field. Axions in
the kink cannot decay into photons because annihilation of the axion would
change the distribution of the axion field in the kink and, hence, increase
the kink's energy. In such a process both the axion field and photon acquire
energy which violates energy conservation.

\section{The alternative theory of gravity vs Einstein general relativity}

So far Einstein general relativity has successfully passed all available
tests. However such tests have inspected the theory only at weak
gravitational field \cite{Will06}. One should note that observations of
binary pulsars yet have not provided a test of general relativity at strong
gravity. Rather, such observations tested Einstein equations in the
post-Newtonian approximation and the strong equivalence principle \cite%
{Will06}.

Are Einstein equations valid for strong gravity? The answer to this question
remains unknown and only appropriate observational tests can shed light on
it. It is well known that in Einstein general relativity the gravitational
field disobeys the principle of superposition. This is the consequence of
the postulate that the space-time metric is determined by the Einstein
equations 
\begin{equation}
R_{ik}-\frac{1}{2}g_{ik}R=8\pi T_{ik},  \label{ee}
\end{equation}%
with the particular choice of the matter energy-momentum tensor $T_{ik}$
proposed by Einstein. However the Einstein theory can be modified by
modifying the energy-momentum tensor. This yields a possibility to satisfy
the superposition principle by a proper choice of $T_{ik}$.

In Appendix A we derive a space-time metric produced by a static mass
distribution based only on the superposition principle. The answer is given
by the Yilmaz exponential metric. Then we show that the Einstein equations
yield the exponential metric if $T_{ik}$ is taken as an \textquotedblleft
electrostatic\textquotedblright\ energy-momentum tensor.

In the weak field limit the exponential metric is equivalent to those
obtained in Einstein theory and, hence, the exponential metric agrees with
the four classic tests of general relativity. However in the opposite limit
of strong gravity the exponential metric is dramatically different. Since
the superposition principle is satisfied the exponential metric predicts no
black holes, but rather stable compact objects with no event horizon and
very large, but finite, gravitational redshift (\textquotedblleft dark red
holes"). This suggests that gravitational field for those objects
effectively becomes repulsive at large gravitational potential.

Here we analyze properties of compact objects at galactic centers and show
that they are in favour of the exponential metric. Our conclusion is based
on a quantitative analysis which is independent of the particular choice of
the time-dependent theory of gravity \cite{pop}. This is possible because
the main part of the bubble dynamics we use for the quantitative comparison
occurs in the well-tested limit of Newtonian gravity. Only a small part of
the trajectory near the lower radius turning point (where gravity
effectively becomes repulsive) is beyond Newtonian description. As a result,
e.g., the period of bubble oscillation can be accurately obtained using
Newtonian gravity, independent of which theory of gravity yields the
repulsive force at small radius.

\subsection{Bubbles in exponential metric}

In Appendix A we derive a metric for a static mass distribution assuming
that the gravitational field obeys the principle of superposition for any
field strength. The answer is given by an exponential isotropic line element
of the class proposed by Yilmaz \cite{Yilm58,Yilm77}

\begin{equation}
ds^{2}=-e^{2\phi }dt^{2}+e^{-2\phi }(dx^{2}+dy^{2}+dz^{2}),  \label{ymet}
\end{equation}%
where for a spherically symmetric thin-wall bubble of radius $R$ (units are $%
c=G=1$) 
\begin{equation}
\phi (r)=\left\{ 
\begin{array}{c}
-M/r,\quad r\eqslantgtr R \\ 
-M/R,\quad r<R,%
\end{array}%
\right.  \label{ymet1}
\end{equation}%
and $M$ is the bubble mass. The exponential metric does not have the
singularity of the Schwarzschild solution at finite radius, and therefore
replaces the concept of black holes with that of \textquotedblleft dark red
holes".

In the reference frame of a distant observer the energy of a static bubble
is given by \cite{Clap73} 
\begin{equation}
U(R)=4\pi \sigma R^{2}\exp \left( \frac{M}{R}\right) ,  \label{y1}
\end{equation}%
where $\sigma $ is the intrinsic surface energy density (as it would appear
to an observer located at the bubble surface) given by Eq. (\ref{y0}), $M$
is the dimensionless bubble mass in units of $m_{\text{pl}}^{2}/m$ and $R$
is the dimensionless bubble radius in units of $r_{0}$. If the radius of a
spherical bubble changes with time then the total energy is%
\begin{equation}
E=U(R)+E_{k},  \label{y2}
\end{equation}%
where $E_{k}\geq 0$ is the kinetic energy. The total energy $E$ is a
constant of motion, then based on the equivalence principle we obtain $M=E=$%
const. Evolution of the bubble radius is similar to a one dimensional motion
of a particle in the effective potential $U(R)$. We plot $U(R)$ in Fig. \ref%
{u1}. The effective potential has a shape of a well and depends on the total
energy (mass). At $R\gg M$ one can omit gravity and $U(R)\simeq 4\pi \sigma
R^{2}$ is just a surface energy (tension) which tends to contract the
bubble. At $R\ll M$ gravity produces large repulsive effective potential
which forces the bubble to expand. As a result the bubble radius $R(t)$
oscillates between two turning points determined by $U(R)=M$.

Fig. \ref{tp} shows numerical solution of the equation for turning points $%
M=4\pi \sigma R^{2}\exp \left( M/R\right) $. For $4\pi \sigma
M<4/e^{2}=0.541 $ equation has two solutions; radius of such bubbles
oscillates with time between turning points $R_{1}$ and $R_{0}$. If $M=M_{%
\text{max}}=1/(e^{2}\pi \sigma )$ the bubble is static with $R=M/2$. Such
static bubble possesses maximum possible mass. Bubbles with $M>M_{\text{max}%
} $ do not exist. As we show below, for the Galactic center bubble $M\lll M_{%
\text{max}}$.

\begin{figure}[tbp]
\bigskip 
\centerline{\epsfxsize=0.45\textwidth\epsfysize=0.35\textwidth
\epsfbox{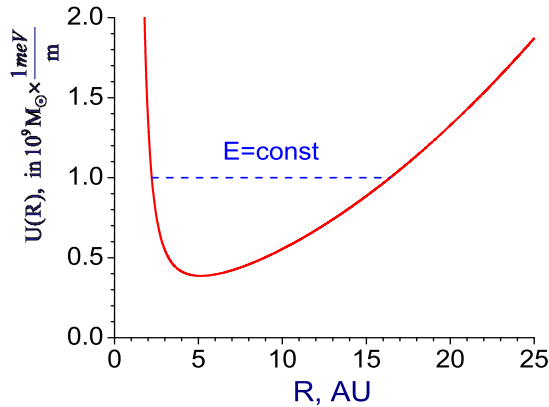}}
\caption{Effective potential for axion bubble motion $U(R)$ (solid line) in
Yilmaz exponential metric. Bubble radius $R(t)$ oscillates between turning
points determined by the equation $U(R)=E$. We expressed $R$ in Astronomical
Units (AU) and $U(R)$ in solar masses, $m$ is the axion mass. }
\label{u1}
\end{figure}

\begin{figure}[tbp]
\bigskip 
\centerline{\epsfxsize=0.45\textwidth\epsfysize=0.35\textwidth
\epsfbox{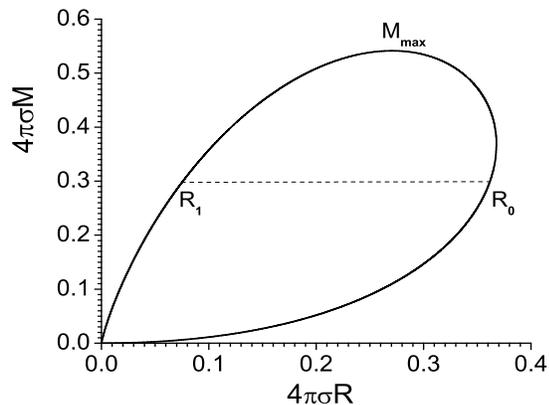}}
\caption{Mass-radius relation for a bubble in Yilmaz exponential metric. At
a given mass $M$ the bubble radius $R(t)$ oscillates between turning points $%
R_1$ and $R_0$. If $M=M_{max}$ then $R_1=R_0$ and the bubble is static. }
\label{tp}
\end{figure}

In a general case to describe $R(t)$ quantitatively we need to use dynamic
equations. These equations depend on a particular choice of the
time-dependent theory of gravity. In this paper, however, we do not need
them which makes our results quite general. The point is that if the
Galactic center object is an axion bubble the main part of its periodic
oscillation occurs in the limit $R(t)\gg M$. In this region we can omit
gravity and use the well-tested special theory of relativity that yields the
following mass-radius equation for a relativistic bubble 
\begin{equation}
M=\frac{4\pi \sigma R^{2}}{\sqrt{1-(dR/dt)^{2}}}
\end{equation}%
which has a simple solution%
\begin{equation}
R(t)=R_{0}\text{cn}\left( \frac{\sqrt{2}t}{R_{0}},\frac{1}{\sqrt{2}}\right) ,
\label{y3}
\end{equation}%
where $R_{0}=\sqrt{M/4\pi \sigma }$ is the maximum bubble radius and cn$%
(x,k) $ is Jacobian elliptic cosine. For small $R(t)$ the solution (\ref{y3}%
) is not applicable. In this region the bubble shrinking slows down and
after reaching the inner turning point $R_{1}$ the bubble starts to expand.
Assuming that Eq. (\ref{y3}) is accurate for the main part of the motion we
obtain that the period of bubble oscillation is $T\simeq 2.622R_{0}$. Taking
into account that $M=4\pi \sigma R_{0}^{2}=8R_{0}^{2}/\alpha ^{2}$, we get
in dimension units $T\simeq 0.262\sqrt{M/m}\hbar /fc^{2}$. Then using Eq. (%
\ref{f1}) we find%
\begin{equation}
T\simeq 0.523\frac{\hbar }{c^{2}}\frac{\sqrt{Mm}}{m_{\pi }f_{\pi }}=\sqrt{%
\frac{M}{10^{6}M_{\odot }}\frac{m}{10^{-3}\text{eV}}}\times 15.27\text{ min}.
\label{y4}
\end{equation}

If $M=3.6\times 10^{6}$M$_{\odot }$ and $T=22.2$ min \cite{Bela06} then Eq. (%
\ref{y4}) yields the axion mass $m\simeq 0.6$ meV ($f\simeq 10^{10}$ GeV).
One should mention, however, that due to time dilation the period of flare
variability depends on the distance between the flare source and the bubble
center (see Fig. \ref{u0}) and, thus, could differ from Eq. (\ref{y4}) by a
factor of the order of one. This yields an inaccuracy in the axion mass
determination in the same factor.

Next we discuss the bubble life time. The bubble decay occurs by means of
axion emission. Due to spherical symmetry there is no radiation of
gravitation waves. Bubble surface, the interface between different vacuum
states, is a soliton (or a kink) that is studied in many areas of nonlinear
physics. One dimension solitons, contrary to 2D or 3D, are stable and
preserve their shape under reflection from a boundary. Because a thin-wall
bubble surface can be treated as a 1D soliton this insures its very long
life time. However, due to finite bubble radius the 1D treatment is only
approximate. Deviation of the problem from 1D leads to slow decay of the
soliton by emission of particles (axions).

We estimate the bubble decay rate as the time of energy loss by the bubble
with the radius $R(t)$ oscillating between the outer $R_{0}$ and inner $%
R_{1} $ turning points. Energy loss by the bubble surface becomes
substantial only when $R(t)\lesssim R_{0}^{2/3}$ (see Appendix B below). In
our case $R_{1}\gg R_{0}^{2/3}$ and, therefore, the region of intensive
energy dissipation is not accessible. As a result, the energy emission is
negligible yielding long-lived bubbles. In Appendix B we estimate the bubble
life-time. The answer is given by Eq. (\ref{s9}) which in dimension units
reads 
\begin{equation}
t\sim \frac{R_{0}}{c}\left( \frac{R_{1}}{R_{0}}\right) ^{4}\left( \frac{R_{1}%
}{r_{0}}\right) ^{2}.  \label{y5}
\end{equation}%
If $m=0.6$ meV then the bubble surface has thickness $r_{0}=\hbar /mc=0.3$
mm. For a bubble with mass $M=3.6\times 10^{6}$M$_{\odot }$ the maximum
radius is $R_{0}=483R_{\odot }$, while the gravitational radius $%
R_{g}=16.1R_{\odot }$. One can find the inner turning point $R_{1}$ from the
equation $R_{0}^{2}=R_{1}^{2}\exp (M/R_{1})$ which yields $R_{1}=1.1R_{\odot
}$. Using Eq. (\ref{y5}) we then obtain the bubble life time $t\sim 5\times
10^{9}$ yrs. This can explain observed lack of supermassive
\textquotedblleft black holes" with $M<10^{6}$M$_{\odot }$. Axion bubbles
with such masses decay fast with life time $t\propto M^{9/2}$.

For bubbles with $M\ll M_{\max }$ one can obtain $R_{1}$ in terms of the
bubble mass $M$ analytically%
\begin{equation}
R_{1}=\frac{M}{2\ln \left[ \frac{2R_{0}}{M}\ln \left( \frac{2R_{0}}{M}%
\right) \right] },\quad R_{0}=\alpha \sqrt{\frac{M}{8}}.
\end{equation}%
Substitute this into Eq. (\ref{y5}) yields the following expression for the
bubble life-time

\[
t\sim \frac{\pi ^{3/2}\hbar m_{\pi }^{3}f_{\pi }^{3}\sqrt{m}M^{9/2}}{\sqrt{8}%
c^{2}m_{\text{pl}}^{12}\ln ^{6}[x\ln (x)]}= 
\]%
\begin{equation}
=\left( \frac{M}{10^{6}M_{\odot }}\right) ^{9/2}\sqrt{\frac{m}{10^{-3}\text{%
eV}}}\times \frac{4.73\times 10^{11}}{\ln ^{6}[x\ln (x)]}\text{ years,}
\label{ylt}
\end{equation}%
where%
\[
x=\frac{m_{\text{pl}}^{2}\sqrt{m}}{\sqrt{2\pi }m_{\pi }f_{\pi }\sqrt{M}}=140%
\sqrt{\frac{m}{10^{-3}\text{eV}}\frac{10^{6}M_{\odot }}{M}}. 
\]%
Eq. (\ref{ylt}) shows that the bubble life-time $t\propto M^{9/2}$ and it
becomes less then the age of the Universe for $M\lesssim 5\times 10^{6}$M$%
_{\odot }$. For $M<10^{6}$M$_{\odot }$ the decay time becomes very short, $%
t\lesssim 10^{7}$ yrs, this is why we do not observe supermassive
\textquotedblleft black holes" with such masses.

Recent analysis of the mass distribution for the compact objects at galactic
centers shows existence of an upper limit for the supermassive
\textquotedblleft black hole" mass \cite{Wu02}:%
\begin{equation}
M_{\max }=1.2_{-0.4}^{+2.6}\times 10^{9}M_{\odot }.  \label{ymass}
\end{equation}%
Next we calculate the maximum possible mass of an axion bubble. In dimension
units it is given by%
\begin{equation}
M_{\max }=\frac{0.0215m_{\text{pl}}^{4}m}{m_{\pi }^{2}f_{\pi }^{2}}.
\label{y6}
\end{equation}%
For $m=0.6$ meV Eq. (\ref{y6}) yields $M_{\max }=1.5\times 10^{9}$M$_{\odot
} $. This value agrees with the upper limit on the supermassive
\textquotedblleft black hole" mass (\ref{ymass}) measured for galactic
nuclei. Radius of the static bubble with $M_{\max }$ is $R=1673R_{\odot }$.

\subsection{Bubble in Einstein general relativity}

\begin{figure}[tbp]
\bigskip 
\centerline{\epsfxsize=0.45\textwidth\epsfysize=0.35\textwidth
\epsfbox{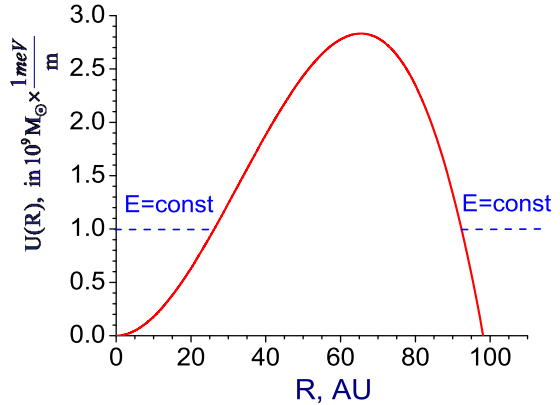}}
\caption{Effective potential for axion bubble motion in Einstein general
relativity. ``Small" bubbles shrink to black holes, while large bubbles
expand infinitely. We expressed $R$ in Astronomical Units (AU) and $U(R)$ in
solar masses.}
\label{ue}
\end{figure}

In Einstein general relativity for a spherically symmetric bubble the metric
can be written in the form

\begin{equation}
ds^{2}=-hdt^{2}+gdr^{2}+r^{2}d\Omega ^{2},
\end{equation}%
where $g$, the radial metric, and $h$, the lapse, are functions of $t$ and $%
r $ with $r$ being the circumferential radius. For a static thin-wall bubble
of radius $R$ the Einstein equations yield 
\begin{equation}
h(r)=\left\{ 
\begin{array}{c}
1-2M/r,\quad r\eqslantgtr R \\ 
1-2M/R,\quad r<R,%
\end{array}%
\right.
\end{equation}%
\begin{equation}
g(r)=\left\{ 
\begin{array}{c}
\frac{1}{1-2M/r},\quad r>R \\ 
1,\quad r<R.%
\end{array}%
\right.
\end{equation}%
Note that $g(r)$ undergoes a jump at the bubble surface, while exponential
metric (\ref{ymet}), (\ref{ymet1}) is continuos. Energy of the thin-wall
bubble with radius $R(t)$ is given by \cite{Blau87,Auri91} 
\begin{equation}
E=\frac{4\pi \sigma R^{2}}{\sqrt{1-(dR/d\tau )^{2}}}-8\pi ^{2}\sigma
^{2}R^{3},
\end{equation}%
where $\tau $ is the interior coordinate time ($d\tau ^{2}=hdt^{2}$). The
corresponding effective potential%
\begin{equation}
U(R)=4\pi \sigma R^{2}-8\pi ^{2}\sigma ^{2}R^{3}
\end{equation}%
is pictured in Fig. \ref{ue}.

In Einstein theory \textquotedblleft small" bubbles shrink toward the
gravitation radius $R_{g}=2M$ and at $t\gg R_{g}/c$ behave as black holes,
while large bubbles expand infinitely. This is dramatically different from
bubble evolution in the exponential metric which we discussed in the
previous section. At the same time, the exponential and the isotropic form
of the Schwarzschild metric (\ref{a2}) are the same to second order in the
gravitational potential $\phi $ in the temporal part and to first order in
the spatial part (see Appendix A). This is sufficient to insure that they
both give identical results in the four classic weak-field tests of general
relativity.

\section{Discussion}

If axion bubbles, rather then supermassive black holes, are located at
galactic centers then what is the mechanism of their nucleation? Dark matter
axions, if they exist, form halos around galaxies. The halo of axions is in
a quantum degenerate non-equilibrium regime. Evolution of the axion halo is
governed by the self-gravity and axion interaction $V(\varphi )$. The
interaction $V(\varphi )$ becomes important only for dense axion clumps.
Dynamics of a dilute galactic halo is determined by self-gravity. Seidel and
Suen have studied evolution of a massive, self-gravitating real scalar field
in Newtonian limit (omitting self-interaction $V(\varphi )$) \cite{Seid94}.
They have shown that independent of the initial conditions a scalar field
configuration collapses to form a compact object by ejecting part of the
scalar field, carrying out the excess kinetic energy. The cooling occurs due
to nonlinear effects of the self-gravitation of the field. Characteristic
cooling time is a free falling time to the center due to self gravity $%
t\simeq 2R_{\text{halo}}^{3/2}/\sqrt{2GM_{\text{halo}}}$, where $R_{\text{%
halo}}$ and $M_{\text{halo}}$ is an initial radius and mass of the axion
halo in a galaxy. For $R_{\text{halo}}=60$ kpc and $M_{\text{halo}}=10^{12}$M%
$_{\odot }$ we obtain the characteristic cooling time $t\sim 10^{8}$ yrs.

Thus, within about $10^{8}$ yrs the gravitational cooling mechanism yields
formation of compact axion clumps in a galactic halo. Evolution of such
clumps is then governed by self-interaction $V(\varphi )$ which leads to
bubble formation. This is shown by three-dimensional numerical simulation of
the evolution of inhomogeneities in the axion field due to the
self-interaction $V(\varphi )$ \cite{Kolb94}. Such a simulation (which omits
gravity) has indeed demonstrated formation of bubble-like structures (see
Fig. 5a in Ref. \cite{Kolb94}). The mass of the nucleated bubbles is much
smaller then the mass of the halo they are born in. However for typical
halos many bubbles are born with masses much greater then $10^{6}$M$_{\odot
} $ and hence they are long-lived objects.

In Einstein general relativity axion bubbles under the influence of surface
tension collapse fast into black holes. However, the Einstein theory yet to
be tested in the limit of strong gravitational field. There is a possibility
that at strong field the gravity is not described by Einstein general
relativity, and rather by an alternative theory which also passes all
available tests. In this paper we consider axion bubbles in a very general
approach avoiding a particular choice of the alternative time-dependent
theory of gravity. Our results are valid for any metric theory which in
based on the principle of superposition. This principle yields the
exponential metric in the static limit, as shown in Appendix A. One should
mention that if the space-time geometry is described by the exponential
metric then compact supermassive objects at galactic centers can not be made
of baryonic matter. Maximum mass of a compact (neutron star like) baryonic
object in such a metric can not exceed about $12$M$_{\odot }$ \cite{Robe99}.
Hence, dark matter of non baryonic origin is the only alternative for their
composition.

We found that in the exponential metric the axion bubbles with $M>10^{6}$M$%
_{\odot }$ are very long lived. Instead of collapsing into a black hole the
bubble radius oscillates between two turning points determined by the net
mass. Such oscillating bubbles, rather then supermassive black holes, could
be present at galactic centers. Our result can account for periodic
variability observed in near-infrared and X-ray flares from Sagittarius A* 
\cite{Genz03,Gill06,Bela06} and yields the axion mass about $0.6$ meV.

Moreover, the bubble scenario with no free parameters (if we fix $m=0.6$ meV
based on Sagittarius A* flare variability) explains lack of supermassive
\textquotedblleft black holes" with $M<10^{6}$M$_{\odot }$. We find that if $%
M<10^{6}$M$_{\odot }$ the bubble life time becomes very short, $t\lesssim
10^{7}$ yrs, and as a result such objects are very rare. We also found that
for the exponential metric the bubble mass can not exceed $M_{\max
}=1.5\times 10^{9}$M$_{\odot }$. This, again with no free parameters,
explains the upper limit on the supermassive \textquotedblleft black hole"
mass measured for galactic centers \cite{Wu02}.

For axion with mass $m=0.6$ meV the axion-photon coupling constant is $%
g_{a\gamma }\sim 10^{-13}$GeV$^{-1}$ \cite{Raff07}. Recently it was
argued that the solar corona $X-$ray emission can be explained by solar
axions of the Kaluza-Klein type (that is by axions propagating into extra
dimensions) which are gravitationally trapped by the Sun and decay near the
solar surface \cite{Dile03,Ziou04}. The estimated value of $g_{a\gamma }$
from the analysis of solar corona $X-$rays is similar to our finding; this
is an interesting coincidence. 

Observation of the Galactic center with very long-baseline interferometry
within the next few years will be capable to test theories of gravitation in
the strong field limit. Such an observation will allow us to distinguish
between the black hole (predicted by Einstein general relativity) and the
oscillating axion bubble scenario which we propose in this paper. If future
observations indeed discover periodic appearance of the shadow from the
Galactic center object this will also be a strong evidence for the axion
nature of dark matter and will lead to an accurate measurement of the axion
mass.

I am very grateful to E. Sezgin and N. Suntzeff\ for useful remarks.

\appendix

\section{Derivation of the static exponential metric from the principle of
superposition}

Let us consider a point mass $M$ located at $r=0$. Static gravitational
field produced by the point mass possesses spherical symmetry. Without loss
of generality one can look for the metric in an isotropic form 
\begin{equation}
ds^{2}=-h(r)dt^{2}+g(r)(dr^{2}+r^{2}d\theta ^{2}+r^{2}\sin ^{2}\theta
d\varphi ^{2}).  \label{a1}
\end{equation}%
Einstein equations yield well known Schwarzschild solution \cite{Land95}%
\begin{equation}
h(r)=\left( \frac{1-M/2r}{1+M/2r}\right) ^{2},\quad g(r)=\left( 1+\frac{M}{2r%
}\right) ^{4}.  \label{a2}
\end{equation}

Here we derive $h(r)$ and $g(r)$ in a different way. First we note that a
static gravitational field of any strength has a potential \cite{Land95}%
\begin{equation}
\phi (r)=\ln \sqrt{h(r)}  \label{a3}
\end{equation}%
and the gravitational force acting on a test rest particle with mass $m$ is%
\begin{equation}
\mathbf{f}=-m\nabla \phi \quad \left( f_{\alpha }=-m\frac{\partial \phi }{%
\partial x^{\alpha }}\right) .  \label{a3f}
\end{equation}
In Minkowski space-time the potential $\phi $ satisfies the Poisson equation 
$\Delta \phi =4\pi M\delta (\mathbf{r})$. Writing the Laplacian and the
delta-function in curvilinear coordinates with metric $g_{ik}$ the Poisson
equation yields

\begin{equation}
\frac{1}{\sqrt{-|g_{ik}|}}\frac{\partial }{\partial x^{i}}\left( \sqrt{%
-|g_{ik}|}g^{ik}\frac{\partial \phi }{\partial x^{k}}\right) =\frac{4\pi }{%
\sqrt{-|g_{ik}|}}M\delta (\mathbf{r}),  \label{a4}
\end{equation}%
where $|g_{ik}|$ is determinant of the metric tensor and $g^{ik}$ is the
tensor reciprocal to $g_{ik}$, that is $g_{ik}g^{kl}=\delta _{i}^{l}$. For
the metric (\ref{a1}) Eq. (\ref{a4}) reduces to%
\begin{equation}
\frac{\partial }{\partial r}\left( r^{2}\sqrt{h(r)g(r)}\frac{\partial \phi
(r)}{\partial r}\right) =4\pi M\delta (r).  \label{a5}
\end{equation}%
Eq. (\ref{a5}) (with $\phi $ from Eq. (\ref{a3})) describes a relation
between the functions $h$ and $g$ which the metric must satisfy. We note
that Eq. (\ref{a5}) is consistent with Einstein general relativity because
the Schwarzschild solution (\ref{a2}) obeys Eq. (\ref{a5}).

To find the functions $h$ and $g$ we need an additional constraint. Here we
postulate that the force of gravity must obey the principle of superposition
at any strength of the gravitational field. This is the only difference from
Einstein general relativity we introduce. Such way of thoughts makes an
appealing connection with the quantum theory.

The principle of superposition is formulated in coordinate systems in which
we measure space coordinates by ideal rods unaffected by gravity (e.g. of
atomic constitution). This assures that the coordinate system is independent
of the position and the value of masses. Let us consider masses $M_{1}$ and $%
M_{2}$ located at coordinates $\mathbf{r}_{1}$ and $\mathbf{r}_{2}$ in the
mentioned above coordinate system. The superposition principle means that
the gravitational force on any test particle due to masses $M_{1}$ and $%
M_{2} $ equals to the vector sum of the force due to the mass $M_{1}$
located at $\mathbf{r}_{1}$ if there is no mass $M_{2}$ and the force due to
the mass $M_{2}$ located at $\mathbf{r}_{2}$ if there is no mass $M_{1}$.

Please note that we formulate the superposition principle for the covariant
force vector $f_{\alpha }$ as in Eq. (\ref{a3f}). This makes a proper
connection with the Newtonian limit in which the measurable gravitational
force is a covariant vector (under coordinate transformation it transforms
like derivatives of a scalar). The covariant force $f_{\alpha }$ on a test
particle is defined as \cite{Land95} $f_{\alpha }=md^{2}x_{\alpha }/ds^{2}$,
where $x_{\alpha }$ is the coordinate of the test particle with mass $m$ and 
$ds$ is the interval. If at the moment when the force is measured the
particle has zero velocity (that is when Eq. (\ref{a3f}) applies) then $%
ds=d\tau $, where $\tau $ is the proper time and, hence, $f_{\alpha
}=md^{2}x_{\alpha }/d\tau ^{2}$. This equation shows that the definition of
the covariant force is unaffected by gravity (if the coordinates $x_{\alpha
} $ are unaffected) and therefore $f_{\alpha }$ is the relevant vector to
formulate the superposition principle. On the other hand, the contravariant
vector $f^{\alpha }=g^{\alpha \beta }f_{\beta }$ contains metric in its
definition and hence cannot be used in the superposition principle.

It follows from Eqs. (\ref{a3f}) and (\ref{a5}) that the force of gravity
satisfies the principle of superposition if and only if 
\begin{equation}
h(r)g(r)=1,  \label{a19}
\end{equation}%
which yields for arbitrary field strength%
\begin{equation}
\frac{\partial }{\partial r}\left( r^{2}\frac{\partial \phi (r)}{\partial r}%
\right) =4\pi M\delta (r).
\end{equation}%
Hence $\phi (r)=-M/r$ and 
\begin{equation}
h(r)=\exp (-2M/r),\quad g(r)=\exp (2M/r).  \label{a20}
\end{equation}%
Eq. (\ref{a20}) is known as Yilmaz exponential metric \cite{Yilm58}.

For small $M/r$ both the Schwarzschild (\ref{a2}) and the exponential (\ref%
{a20}) metrics yield the same expansion%
\begin{equation}
h(r)=1-\frac{2M}{r}+\frac{2M^{2}}{r^{2}}+\ldots ,\quad g(r)=1+\frac{2M}{r}%
+\ldots  \label{a21}
\end{equation}%
and hence both metrics pass the four classic weak-field tests of general
relativity. Accuracy of current tests is yet far from ability to check the
next terms in the expansion (\ref{a21}) where the two metrics start to
deviate from each other \cite{Will06}.

The principle of superposition allows us to find the metric in the case of $%
N $ point masses $M_{1}$, ..., $M_{N}$ located at $\mathbf{r}_{1}$, ... $%
\mathbf{r}_{N}$. To apply the principle of superposition we must choose the
coordinate system in which we measure space coordinates by ideal rods
unaffected by gravity. Since the speed of light measured by such rods and
clocks (e.g. of atomic constitution) is independent of direction in a
gravitational field the metric in such coordinates is isotropic, that is 
\begin{equation}
ds^{2}=-h(\mathbf{r})dt^{2}+g(\mathbf{r})(dx^{2}+dy^{2}+dz^{2}).
\label{a21a}
\end{equation}%
For the metric (\ref{a21a}) Eq. (\ref{a4}) yields%
\[
\frac{\partial }{\partial x}\left( \sqrt{hg}\frac{\partial \phi }{\partial x}%
\right) +\frac{\partial }{\partial y}\left( \sqrt{hg}\frac{\partial \phi }{%
\partial y}\right) +\frac{\partial }{\partial z}\left( \sqrt{hg}\frac{%
\partial \phi }{\partial z}\right) = 
\]%
\begin{equation}
=4\pi \lbrack M_{1}\delta (\mathbf{r}-\mathbf{r}_{1})+\ldots +M_{N}\delta (%
\mathbf{r}-\mathbf{r}_{N})].  \label{a21b}
\end{equation}%
The principle of superposition for the potential $\phi $ (and hence for the
force of gravity $\mathbf{f}$) is satisfied provided $hg=1$. Then Eqs. (\ref%
{a21b}) and (\ref{a3}) give the following answer for $N-$body space-time
geometry in isotropic Cartesian coordinates 
\begin{equation}
ds^{2}=-e^{2\phi }dt^{2}+e^{-2\phi }(dx^{2}+dy^{2}+dz^{2}),  \label{a22}
\end{equation}%
where $\phi $ is the $N-$body potential%
\begin{equation}
\phi (\mathbf{r})=-\frac{M_{1}}{|\mathbf{r}-\mathbf{r}_{1}|}-...-\frac{M_{N}%
}{|\mathbf{r}-\mathbf{r}_{N}|}.  \label{a23}
\end{equation}%
Thus, finding the space-time geometry reduces to a simple \textquotedblleft
electrostatic" problem.

Next we discuss how to obtain the exponential metric from Einstein equations.

\subsection{Derivation of exponential metric from Einstein equations}

Einstein equations read%
\begin{equation}
R_{ik}-\frac{1}{2}g_{ik}R=8\pi T_{ik},  \label{a24}
\end{equation}%
where $R_{ik}$ is the Ricci tensor, $R$ is the scalar space curvature and $%
T_{ik}$ is the energy-momentum tensor of matter. Let us consider $N$ fixed
point masses $M_{1}$, ..., $M_{N}$ located at $\mathbf{r}_{1}$, ... $\mathbf{%
r}_{N}$. \ For such a system the energy-momentum tensor in Einstein general
relativity is%
\begin{equation}
T_{0}^{0}=\sum_{j=1}^{N}M_{j}\delta (\mathbf{r}-\mathbf{r}_{j}),\quad \text{%
all other }T_{i}^{k}=0.  \label{a25}
\end{equation}%
Solution of Eq. (\ref{a24}) with this $T_{ik}$ yields black holes and no
superposition principle.

To obtain exponential metric we must use another energy-momentum tensor. Let
us write $T_{ik}$ by analogy with electrostatic. For $N$ fixed point
electric charges $q_{1}$, ..., $q_{N}$ located at $\mathbf{r}_{1}$, ... $%
\mathbf{r}_{N}$ the energy-momentum tensor is given by \cite{Land95} (in
curvilinear coordinates)%
\begin{equation}
T_{ik}=\frac{1}{4\pi }\left( \frac{\partial \phi }{\partial x^{i}}\frac{%
\partial \phi }{\partial x^{k}}-\frac{1}{2}g_{ik}(\nabla \phi )^{2}\right) ,
\label{a26}
\end{equation}%
where $(\nabla \phi )^{2}=E_{\alpha }E^{\alpha }=g^{\alpha \beta }(\partial
\phi /\partial x^{\alpha })(\partial \phi /\partial x^{\beta })$ and $\phi $
is the electric potential satisfying the Poisson equation%
\begin{equation}
\frac{\partial }{\partial x^{i}}\left( \sqrt{-|g_{ik}|}g^{ik}\frac{\partial
\phi }{\partial x^{k}}\right) =4\pi \sum_{j=1}^{N}q_{j}\delta (\mathbf{r}-%
\mathbf{r}_{j}).  \label{a27}
\end{equation}%
We assume that for the system of $N$ fixed point masses $M_{1}$, ..., $M_{N}$
the energy-momentum tensor $T_{ik}$ is given by Eqs. (\ref{a26}) and (\ref%
{a27}) with the change $q_{j}\rightarrow M_{j}$. Substituting this tensor
into Einstein equations (\ref{a24}) we obtain the solution for the metric
given by formulas (\ref{a22}) and (\ref{a23}).

One should mention that we can find the proper energy-momentum tensor (\ref%
{a26}) simply by plugging the exponential metric (\ref{a22}) into the left
hand side of Einstein equations (\ref{a24}). Then the \textquotedblleft
electrostatic" energy-momentum tensor (\ref{a26}) is obtained automatically.

The result discussed here is valid for a static gravitational field. How to
generalize it for time-dependent fields is beyond the scope of the present
paper.

\begin{widetext}
\onecolumngrid

\section{Energy emission from a shrinking bubble}

Here we calculate energy loss by a shrinking spherically symmetric bubble
caused by emission of scalar particles (axions). For an order of magnitude estimate
one can omit the effect of gravity. Then the evolution of the scalar field $%
\varphi (t,r)$ is described by sine-Gordon equation
\begin{equation}
\ddot{\varphi}-\varphi ^{\prime \prime }+\frac{1}{\alpha }\sin (\alpha
\varphi )=2\varphi ^{\prime }/r,  \label{s1}
\end{equation}%
where $r$ is the radial coordinate. Without right-hand side, Eq. (\ref{s1})
has an exact, so-called kink, solution
\begin{equation}
\varphi _{0}=\frac{4}{\alpha }\arctan \left\{ \exp \left[ \pm \frac{%
(r-vt-R_{0})}{\sqrt{1-v^{2}}}\right] \right\} ,  \label{s2}
\end{equation}%
where $R_{0}\gg 1$ is the initial bubble radius. The solution describes a
kink (space region where $\varphi $ changes from $2\pi /\alpha $ to $0$)
propagating with constant velocity $v$; the kink's size is $l\sim \sqrt{%
1-v^{2}}$.

If $l\ll R(t)$, where $R(t)$ is the bubble radius, r.h.s. of (\ref{s1}) may
be treated as a small perturbation. Eq. (\ref{s1}) possesses approximate
solution in the form of the kink (\ref{s2}) with parameters slowly changing
in time under the action of the perturbation. In particular, the kink
shrinks due to its surface tension so that the bubble radius and the
velocity evolve as \cite{Malo87a}
\begin{equation}
R(t)=R_0cn(\sqrt{2}t/R_0,1/\sqrt{2})\text{, \ }v(t)=\sqrt{1-R^4(t)/R_0^4}%
\text{,}
\end{equation}
where $cn$ stands for the elliptic cosine with the modulus $1/\sqrt{2}$.
Such a process is accompanied by emission of scalar particles which yields
the energy loss. We estimate the energy loss following the original work of
Malomed \cite{Malo87a,Malo87b}. In terms of the inverse scattering
technique, the spectral density of the emitted energy $E_e(t,q)$ is

\begin{equation}
\frac{dE_{e}}{dq}=\frac{4}{\pi \alpha ^{2}}|B(t,q)|^{2},  \label{s3}
\end{equation}%
where $q$ is the radiation wavenumber and the perturbation-induced evolution
equation for the complex amplitude $B(t,q)$ is given by \cite%
{Malo87a,Malo87b}
\begin{equation}
\frac{dB}{dt}=-\frac{i}{2(\lambda ^{2}+\gamma ^{2})}\int_{-\infty }^{\infty
}dr\left( \lambda ^{2}-\gamma ^{2}-2i\lambda \gamma \tanh \left[ \frac{r-vt}{%
\sqrt{1-v^{2}}}\right] \right) \exp \left( i\sqrt{1+q^{2}}t-iqr\right)
\partial _{r}\varphi _{0},  \label{s4}
\end{equation}%
where $\lambda =\sqrt{1+q^{2}}-q$ and $\gamma =(1+v)/2\sqrt{1-v^{2}}$.
Calculating the integral in (\ref{s4}) yields
\begin{equation}
\frac{dB}{dt}=\frac{i\pi \left[ \lambda ^{2}(1-v)(1-\sqrt{1+v})-v/2\right] }{%
(1+v)/4+\lambda ^{2}(1-v)}\frac{\exp \left( i\sqrt{1+q^{2}}t-iqvt\right) }{%
\cosh \left[ \pi q\sqrt{1-v^{2}}/2\right] }.  \label{s5}
\end{equation}%
If $v$ slowly varies with time one can take $v\approx const$ in Eq. (\ref{s5}%
), then after integration we obtain
\begin{equation}
B(t,q)=\frac{\pi \left[ \lambda ^{2}(1-v)(1-\sqrt{1+v})-v/2\right] }{\left[
(1+v)/4+\lambda ^{2}(1-v)\right] (\sqrt{1+q^{2}}-qv)}\frac{\exp \left( i%
\sqrt{1+q^{2}}t-iqvt\right) -1}{\cosh \left[ \pi q\sqrt{1-v^{2}}/2\right] }.
\end{equation}%
Therefore
\begin{equation}
\frac{dE_{e}}{dq}=\frac{16\pi \left[ \lambda ^{2}(1-v)(1-\sqrt{1+v})-v/2%
\right] ^{2}}{\alpha ^{2}\left[ (1+v)/4+\lambda ^{2}(1-v)\right] ^{2}(\sqrt{%
1+q^{2}}-qv)^{2}}\frac{\sin ^{2}\left[ \left( \sqrt{1+q^{2}}-qv\right) t/2%
\right] }{\cosh ^{2}\left[ \pi q\sqrt{1-v^{2}}/2\right] }.  \label{s6}
\end{equation}%
Integration of (\ref{s6}) over $dq$ gives the emitted energy as a function
of time $E_{e}(t)=\int_{-\infty }^{\infty }dq(dE_{e}/dq)$. In Eq. (\ref{s6})
sine is a fast oscillating function, so we substitute $\sin
^{2}(x)\rightarrow 1/2$. The radiation power increases when the kink's
velocity $v$ approaches the speed of light $c=1$. Assuming $1-v\ll 1$,
integration of Eq. (\ref{s6}) yields
\begin{equation}
E_{e}(t)\approx \frac{2.51}{\alpha ^{2}(1-v(t))^{3/2}}\approx \frac{7.10}{%
\alpha ^{2}}\left( \frac{R_{0}}{R(t)}\right) ^{6}  \label{s7}
\end{equation}%
The emitted energy becomes comparable with the initial bubble energy $%
E_{0}=4\pi \sigma R_{0}^{2}=8R_{0}^{2}/\alpha ^{2}$ when the bubble radius
reaches the value $R_{\ast }\approx R_{0}^{2/3}$. This value agrees with
those obtained in \cite{Widr89}.

If the bubble shrinks from the outer turning point $R_{0}$ to the inner
turning point $R_{1}\gg R_{\ast }$ the radiated energy is
\begin{equation}
E_{e}\sim \frac{7.10}{\alpha ^{2}}\left( \frac{R_{0}}{R_{1}}\right) ^{6}%
\text{.}  \label{s8}
\end{equation}%
To emit all its energy the bubble must oscillate between $R_{1}$ and $R_{0}$
about $E_{0}/E_{e}$ cycles. As a result, the bubble life-time is
\begin{equation}
t\sim R_{0}\frac{E_{0}}{E_{e}}\approx \frac{R_{1}^{6}}{R_{0}^{3}}.
\label{s9}
\end{equation}

\end{widetext}\twocolumngrid

\end{document}